\documentstyle[epsf]{l-aa}
\begin{document}
\thesaurus{20(04.03.1	% Catalogs
	   11.01.2	% Galaxies: active
	   11.17.3	% Quasars: general
	   13.18.1)}	% Radio continuum: general
\title{Gigahertz Peaked Spectrum sources from the \\ Jodrell Bank--VLA 
Astrometric Survey}
\subtitle{I. Sources in the region $35\degr \leq \delta \leq 75\degr$}
\author{A. Marecki\inst{1}
        \and H. Falcke\inst{2}
        \and J. Niezgoda\inst{1}
        \and S. T. Garrington\inst{3}
	\and A. R. Patnaik\inst{2}}
\offprints{Heino Falcke, email: hfalcke@mpifr-bonn.mpg.de}
\institute{Centre for Astronomy, N. Copernicus University, ul. Gagarina 11,
           PL-87-100 Toru\'n, Poland
           \and
           Max-Planck-Institut f\"ur Radioastronomie,
           Auf dem H\"ugel 69, D-53121 Bonn, Germany
           \and
           Nuffield Radio Astronomy Laboratories, Jodrell Bank,
           Cheshire, SK11 9DL, UK}

\date{Received 13 May 1998; accepted 13 October 1998}

\maketitle

\begin{abstract}

Observations with MERLIN\footnote{
MERLIN is operated by The University of Manchester on behalf of the UK Particle
Physics and Astronomy Research Council.} at 408~MHz have been used to establish
the low-frequency part of the spectra of more than a hundred compact
radio sources taken from the part of the Jodrell Bank--VLA Astrometric
Survey limited by $35\degr \leq \delta \leq 75\degr$. These sources
were selected from JVAS and other catalogues to have convex spectra
between 1.4 and 8.4~GHz, characteristic of Gigahertz Peaked Spectrum
(GPS) sources.  We have confirmed convex shapes of the spectra of
76~objects (one half of our initial candidates) thereby yielding the
largest genu\-ine sample of GPS sources compiled so far. Seven of
17~identified quasars in the sample have large ($z\ga 2$) redshifts.
 
\end{abstract}

\section{Introduction}

By definition, Gigahertz Peaked Spectrum (GPS) sources have convex spectra
with a turnover frequency $\nu_{max}>1$~GHz. It is generally accepted that
such a spectral shape results from synchrotron self-absorption due to a
high compactness. Indeed, the linear sizes of GPS sources are very small
(10--1000~pc) and their radio luminosities are large ($L_{radio}\sim10^{45}$
erg/s). An important feature which makes the GPS class particularly
interesting, is that GPS objects identified with quasars often have very
large redshifts. A review on the properties of GPS sources has been given
by O'Dea et al.~(1991) --- hereinafter OBS91 --- and O'Dea~(1998). De Vries
et al.~(1997) compiled spectra of 72~GPS sources and constructed a canonical
GPS radio spectrum. They found it to have a
constant shape independent of AGN type, redshift or radio luminosity.

As Gopal-Krishna \&~Spoelstra (1993) pointed out, {\it the great potential
of GPS sources for discovering high-$z$ objects continues to be the major
motivation factor for enlarging the sample of GPS sources}. A second
motivation for increasing the number of known GPS sources comes from the
discovery (Wilkinson et al.~1994) of a new class of Compact Symmetric
Objects (CSO). Five archetypal CSOs (cf. Readhead et al.~1996a) are all
acknowledged GPS sources. Readhead et al. (1996b) argue that CSOs are the
young precursors of classical double radio sources. Increasing the list of
known GPS sources offers a promising way to find more CSOs. 

The search for GPS sources can be difficult because it requires both
high resolution and high sensitivity observations at frequencies well
below 1~GHz and high sensitivity observations at $\nu >\,$ 5~GHz, well
above the potential peak of the spectrum.  Therefore, unlike in the
case of flat spectrum sources, only a modest number of GPS sources can
simply be extracted from classical large catalogues e.g. the Green
Bank (GB) surveys at 1.4~GHz and 4.85~GHz (White \&~Becker~1992 and
references therein --- hereinafter WB92) and the Texas Survey at
365~MHz (Douglas et al.~1996).  For example, the sample of
Stanghellini et al.~(1990) was derived from the Texas Survey and the
1~Jy catalogue of K\"uhr et al.~(1981a). Out of 55~sources and
candidates they listed, 41~are currently acknowledged as GPS sources
(O'Dea, priv. comm.~1996) and 33~make a complete sample (Stanghellini
et al.~1996).

The lists published by Gopal-Krishna et al.~(1983) and Spoelstra et
al.~(1985) resulting from dedicated observations, e.g. with the Westerbork
Synthesis Radio Telescope (WSRT) or the Ooty telescope, are not very large:
each one contains 25~sources (5~and 2~out of these two samples respectively had
been retracted later). The sources gathered by the authors mentioned above
have been collected in OBS91 in the so called "working sample" encompassing
95~objects. Gopal-Krishna \&~Spoelstra~(1993) confirmed the existence of
10~GPS sources and Cersosimo et al.~(1994) found 7~more\footnote{Cersosimo
et al. claim to discover more sources but only 7~have been finally recognised
as GPS.}. On the other hand 6~sources from the list in OBS91, namely
0218+357, 0528+134, 0902+490, 1851+488, 2053$-$201 and 2230+114 proved to be
"not so good" examples of the class and have been retracted. By the end
of 1997 very few other GPS sources had been discovered.

There are two ways to achieve a "bulk" increase of this number. The first
one has been followed by Snellen et al.~(1998) --- hereinafter SSB98 ---
the second is the basis of this paper.
The former one is based on the Westerbork Northern Sky Survey
(WENSS) being carried out with WSRT at 325 and 609~MHz.
Naturally, the sources with inverted spectra in the WENSS
are GPS candidates, which are followed up with observations at higher
frequencies. This survey is particularly useful since it is the most
sensitive survey at low frequencies (more than an
order of magnitude better than the Texas Survey). 

The published part of WENSS called "mini--survey" (Rengelink et
al.~1997) is limited to $14^h 10^m < \alpha < 20^h 30^m$, $57\degr < \delta <
72\degr 50\arcmin$ and covers $\sim570$ square degrees only, so there are
still large parts of the sky where this method cannot be applied.
Nevertheless, SSB98 using the mini--survey plus an unpublished part
of WENSS ($4^h < \alpha < 8^h 30^m$, the same declination range as the
mini--survey) were able to establish a list of 47~"new" GPS sources.

\section{Construction of the sample}

Since WENSS was not available when our programme started in 1994, we
adopted a different approach which can be regarded as "opposite" to that
of SSB98 because we started at the high frequencies end instead of
the low one. Hence, to determine candidate GPS sources, we used 8.4~GHz
fluxes from the first part of the Jodrell Bank--VLA Astrometric Survey
(JVAS)\footnote{JVAS resulted from observations made with the VLA in
"A"~configuration.} (Patnaik et al.~1992) containing all compact radio
sources in the range $35\degr \leq \delta \leq 75\degr$ with fluxes
$>200$~mJy at 5~GHz. We combined them with 1.4~GHz and 4.85~GHz fluxes
from the GB catalogues (WB92). For 32~sources 1.4~GHz fluxes were missing
in WB92 so we substituted them with the catalogue flux density limit. Then
we fitted the data with a second order polynomial of the form:
\begin{equation}
\log S_\nu=S_0+\alpha \log \nu - c (\log \nu)^2,
\end{equation}
where $\nu$ is the observed frequency in GHz and $S_\nu$ the flux
density in mJy and selected the sources with convex spectral shapes.
The criterion for selecting an initial list of
\begin{table}[t]
\caption[ ]{Subsample One}
\begin{flushleft}
\begin{tabular}{rrrc}
\hline\noalign{\smallskip}
IAUname & \multicolumn{1}{c}{R.A.} & \multicolumn{1}{c}{Dec.} & Opt. \\
\multicolumn{1}{c}{(B1950)} & \multicolumn{2}{c}{(J2000)} & ID\\
\noalign{\smallskip}
\hline\noalign{\smallskip}

0059+581 & 01 02 45.7630 & 58 24 11.139 & CW \\
0102+480 & 01 05 49.9295 & 48 19 03.183 & EF \\
0627+532 & 06 31 34.6860 & 53 11 27.754 & BS \\
0652+426 & 06 56 10.6629 & 42 37 02.751 &  G \\
0652+577 & 06 57 12.5027 & 57 41 56.740 & EF \\
0750+535 & 07 54 15.2177 & 53 24 56.450 & EF \\
0828+493 & 08 32 23.2171 & 49 13 21.036 & BL \\
1107+485 & 11 10 36.3237 & 48 17 52.446 & EF \\
1239+552 & 12 41 27.7043 & 54 58 19.040 & EF \\
1256+546 & 12 58 15.6078 & 54 21 52.112 & EF \\
1311+552 & 13 13 37.8518 & 54 58 23.894 & EF \\
1428+422 & 14 30 23.7418 & 42 04 36.503 & EF \\
1532+680 & 15 32 43.3426 & 67 55 13.992 & EF \\
1602+576 & 16 03 55.9311 & 57 30 54.415 & BS \\
1627+476 & 16 28 37.5064 & 47 34 10.414 & BS \\
1630+358 & 16 32 31.2578 & 35 47 37.740 & EF \\
1745+670 & 17 45 54.3577 & 67 03 49.302 & EF \\
1755+578 & 17 56 03.6285 & 57 48 47.990 & BS \\
1815+614 & 18 15 36.7920 & 61 27 11.641 & EF \\
1946+708 & 19 45 53.5197 & 70 55 48.723 &  ? \\
2253+417 & 22 55 36.7082 & 42 02 52.535 & BS \\
2310+385 & 23 12 58.7950 & 38 47 42.668 & BS \\
2323+478 & 23 25 44.9131 & 48 06 25.280 & BS \\
2356+385 & 23 59 33.1809 & 38 50 42.322 & BS \\
\noalign{\smallskip}
\hline
\end{tabular}
\end{flushleft}
\small{
Explanation of ID field:\\

BL: BL Lac object\\
BS: blue stellar object (without the spectrum --- basically a quasar)\\
BG: blue galaxy? \\
CW: crowded field --- too many objects for a clean identification\\
EF: empty field\\
G:  galaxy\\
NS: neutral stellar object (same magnitude in both E and O prints)\\
OB: obscured field\\
Q:  quasar\\
RS: red stellar object \\
}
\end{table}
candidate GPS sources was that the curvature $c$ had to be greater than 1
with a peak in the spectrum in the observable range (the limit of $c>1$
is rather conservative and includes also spectra which are much flatter
than those of typical GPS sources). Admittedly, although this criterion
seems to be conservative enough, a few sources from the OBS91 "working
sample" which happen to be JVAS members, and therefore have been fitted
with polynomials (0018+729, 0248+430, 0552+398, 2015+657, 2021+614,
2352+495), {\it did not} fulfill it.  In the end, the $c>1$ threshold was
chosen as a good compromise, to produce a reasonably sized sample for
future observations. The number of the initial candidates selected with
the above procedure was~163. 

This list contained 14 known GPS sources: 0108+388, 0153+744, 0636+680,
0646+600, 0710+439, 0711+356, 1031+567, 1225+368, 1333+459, 1843+356 and
2050+364 listed in OBS91; 0903+684 from Gopal-Krishna \&~Spoel\-stra~(1993)
and two discovered by Snellen et al.~(1995) --- 0700+470 and 1324+574.
These were not considered for further observations.

We divided the remaining sample of 149~sources into two parts. The first one
--- hereinafter "Subsample One" --- contains those which are found in the
365~MHz Texas survey (24~sources). They are listed in Table~1 along with JVAS
positions rounded to 1~milliarcsecond and our identifications made using POSS.
Nine of these are also listed in the 6C catalogue which has a flux density
limit of 200~mJy at 151~MHz (Hales et al.~1993 and references therein).

For the remaining 125~sources --- hereinafter "Subsample Two" --- we had
virtually no low frequency data. Although 19 of these sources were
present in the B3 survey at 408~MHz (Ficarra et al.~1985), its overlap
in sky coverage with the first part of the JVAS survey we use here is small
and, furthermore, the B3 survey has significant errors in flux density near
the catalogue limit of 100~mJy. 

The aforementioned crude extrapolation of the spectra applied to medium
and high frequency data (1.4~GHz $< \nu < $ 8.4~GHz) of Subsample Two plus
the flux density limit of the Texas Survey gave us only an unreliable estimate
of the frequency turnover and even the convex shape of the spectrum remained
uncertain in some cases. Particularly some weaker sources could either be
GPS or mere flat spectrum depending on whether their 365~MHz fluxes were
far below or {\it just} below the Texas catalogue limit respectively. Another
effect that will produce a spurious convex shape in our non-simultaneous
data is the variability typically observed in flat spectrum sources. 

\section{Observations and data reduction}

In order to investigate the low frequency part of the spectra of Subsample
Two sources we set up a programme of flux density measurements at low
frequency with a high resolution facility. We used MERLIN at 408~MHz
because of its superior resolving capability --- $1\arcsec$. The
observations were carried out in the period from November~1994 until
January~1995 and each source was typically observed twice (with
different hour angles) for about 15~minutes per scan. 

Such short snapshots cannot be used to produce reliable maps with MERLIN,
since the aperture coverage is too sparse. Confusion is a significant
effect at this low frequency and therefore fringe-frequency vs. delay
(FFD) plots were used to separate confusing sources from the central
target source (see eg. Walker~1981). Since the data were not
phase-calibrated the coherence time is limited to a few minutes. By taking
$128\times 4$ second sub-samples of the data and Fourier transforming them
in both time and frequency, a "map" of the field surrounding the source can
be produced whose axes are fringe-frequency and delay. The target source
will have close to zero fringe-frequency and delay, since its position is
well known from the JVAS survey, and appear at the centre of the "map",
while confusing sources will be offset from the centre. The flux density
of the target source was determined from the central peak in the FFD map.
The FFD plots were inspected visually; the detection threshold in a single
plot was set at $3\sigma$ and flux density values were only listed if the
target was detected in two or more plots. In all cases, the values
determined from the FFD plots were consistent with simple averages of
phase-calibrated data on individual baselines calculated using the Astronomical
Image Processing System (AIPS). The flux density scale was
determined using observations of 3C286, for which the Baars et al. (1977)
value was used. 

In Table~2 we report the results of successful measurements of 408~MHz
fluxes for 98~sources from Subsample Two along with their JVAS positions
rounded to 1~milli\-arcsecond and our identifications made using POSS.
Interference and other problems at other sites resulted in the loss of
data for the remaining 27~sources. They are listed in Table~3. The flux
density values given in Table~2 resulted from averaging the measurements
over all baselines involving the Defford or Knockin telescopes (6 or
8~baselines, depending on whether the Lovell telescope was used or not)
and in both LL and RR polarisations. Shorter baselines were not used
because they provide insufficient resolution in delay or fringe-frequency
to separate the confusing sources; additionally the use of longer
baselines reduces any contribution from large-scale halos seen around
some GPS sources. The longest baselines involving the Cambridge telescope
were not used because interference limited the useful bandwidth to
1.5~MHz, rather than the 4~MHz used elsewhere. Because baseline
combinations and the volume of data per source varied form source to
source the errors range form 10 to 30~mJy. 

\begin{table}
\caption[ ]{Subsample Two --- flux densities of sources at 408 MHz}
\begin{flushleft}
\begin{tabular}{rrrrrc}
\hline\noalign{\smallskip}
IAUname & \multicolumn{1}{c}{R.A.} & \multicolumn{1}{c}{Dec.} &
\multicolumn{1}{c}{Flux} & Opt. \\
\multicolumn{1}{c}{(B1950)} & \multicolumn{2}{c}{(J2000)} &
\multicolumn{1}{c}{[mJy]} & ID\\
\noalign{\smallskip}
\hline\noalign{\smallskip}

0001+478 & 00 03 46.0413 & 48 07 04.134 &  104 & BS \\
0015+529 & 00 17 51.7596 & 53 12 19.126 &  100 & BS \\
0046+511 & 00 49 37.9901 & 51 28 13.700 &  123 & NS \\
0051+679 & 00 54 17.6237 & 68 11 11.175 &  127 & OB \\
0051+706 & 00 54 17.6884 & 70 53 56.625 &  260 & NS \\
0058+498 & 01 01 16.9988 & 50 04 44.991 &  125 & BS \\
0102+511 & 01 05 29.5588 & 51 25 46.576 &  262 & NS \\
0123+731 & 01 27 04.7169 & 73 23 12.676 &  330 & EF \\
0129+431 & 01 32 44.1273 & 43 25 32.667 &  244 & BS \\
0129+560 & 01 32 20.4503 & 56 20 40.372 &  113 & CW \\
0140+490 & 01 43 46.8791 & 49 15 41.586 &  105 & CW \\
0148+546 & 01 51 36.2876 & 54 54 37.688 &  100 & EF \\
0153+389 & 01 56 31.4088 & 39 14 30.929 &  119 & BS \\
0213+444 & 02 16 17.1707 & 44 37 43.405 &  154 & EF \\
0251+393 & 02 54 42.6316 & 39 31 34.714 &  114 & BS \\
0307+380 & 03 10 49.8805 & 38 14 53.845 &  148 & NS \\
0335+599 & 03 39 09.3942 & 60 08 56.960 &  134 & RS \\
0336+473 & 03 40 10.7897 & 47 32 27.328 &  103 & BS \\
0338+480 & 03 42 10.3522 & 48 09 46.948 &  113 & NS \\
0412+447 & 04 15 56.5246 & 44 52 49.676 &  110 & OB \\
0424+414 & 04 27 46.0455 & 41 33 01.091 &  272 & EF \\
0454+550 & 04 58 54.8417 & 55 08 42.042 &  192 & BS \\
0513+714 & 05 19 28.8835 & 71 33 03.740 &  153 & EF \\
0514+474 & 05 18 12.0899 & 47 30 55.536 &  220 & CW \\
0533+446 & 05 37 30.0630 & 44 41 03.533 &  155 &  ? \\
0559+422 & 06 02 58.9438 & 42 12 09.999 &  259 & BS \\
0610+510 & 06 14 49.1589 & 51 02 13.124 &  143 & BG \\
0621+446 & 06 25 18.2652 & 44 40 01.628 &  101 & BS \\
0630+497 & 06 33 52.2068 & 49 43 45.939 &  149 & BS \\
0651+410 & 06 55 10.0243 & 41 00 10.148 &  130 &  G \\
0655+696 & 07 01 06.6159 & 69 36 29.414 &  293 & BS \\
0708+742 & 07 14 36.1236 & 74 08 10.142 &  105 & BS \\
0713+669 & 07 18 05.6314 & 66 51 53.332 &  145 & EF \\
0718+374 & 07 22 01.2600 & 37 22 28.628 &   64 & BS \\
0732+755 & 07 39 13.1962 & 75 27 47.702 &  141 & EF \\
0753+373 & 07 56 28.2513 & 37 14 55.647 &   59 & BS \\
0753+519 & 07 56 59.5457 & 51 51 00.237 &   98 & BS \\
0758+594 & 08 02 24.5932 & 59 21 34.800 &  102 & BS \\
0849+675 & 08 53 34.3220 & 67 22 15.665 &  212 & BS \\
0851+719 & 08 56 54.8695 & 71 46 23.894 &  105 & BS \\
0900+520 & 09 03 58.5758 & 51 51 00.658 &  354 &  ? \\
0924+732 & 09 29 42.1565 & 73 04 04.553 &  114 &  G \\
0925+745 & 09 30 53.7823 & 74 20 05.930 &  112 & EF \\
0939+620 & 09 43 14.5025 & 61 50 33.343 &   99 & BS \\
1017+436 & 10 20 27.2021 & 43 20 56.342 &   51 & BS \\
1019+429 & 10 22 13.1324 & 42 39 25.618 &  138 & RS \\
1032+509 & 10 35 06.0176 & 50 40 06.087 &  198 & EF \\
1035+430 & 10 38 18.1899 & 42 44 42.766 &  194 & RS \\
1043+541 & 10 46 24.0372 & 53 54 26.220 &  181 & BS \\
1055+433 & 10 58 02.9208 & 43 04 41.505 &  182 & EF \\
1101+609 & 11 04 53.6946 & 60 38 55.287 &  151 & BS \\
1125+366 & 11 27 58.8707 & 36 20 28.352 &   86 & BS \\
1138+644 & 11 41 12.2283 & 64 10 05.484 &  180 & EF \\
\noalign{\smallskip}
\hline
\end{tabular}
\end{flushleft}
\end{table}
\begin{table}
{\small{\bf Table \thetable{}a.} continued}
\begin{flushleft}
\begin{tabular}{rrrrrc}
\hline\noalign{\smallskip}
IAUname & \multicolumn{1}{c}{R.A.} & \multicolumn{1}{c}{Dec.} &
\multicolumn{1}{c}{Flux} & Opt. \\
\multicolumn{1}{c}{(B1950)} & \multicolumn{2}{c}{(J2000)} &
\multicolumn{1}{c}{[mJy]} & ID\\
\noalign{\smallskip}
\hline\noalign{\smallskip}

1157+532 & 12 00 06.0107 & 53 00 37.118 &  188 &  Q \\
1206+415 & 12 09 22.7884 & 41 19 41.369 &  152 & BS \\
1226+638 & 12 29 06.0256 & 63 35 00.986 &  153 & RS \\
1232+366 & 12 35 05.8076 & 36 21 19.308 &  121 & BS \\
1239+606 & 12 41 29.5907 & 60 20 41.320 &  136 & BS \\
1245+676 & 12 47 33.3300 & 67 23 16.457 &  107 &  G \\
1300+485 & 13 02 17.1974 & 48 19 17.572 &  163 & BS \\
1308+471 & 13 10 53.5906 & 46 53 52.219 &  160 & EF \\
1320+394 & 13 22 55.6615 & 39 12 07.984 &  119 & BS \\
1321+410 & 13 24 12.0940 & 40 48 11.773 &   92 & BS \\
1337+637 & 13 39 23.7812 & 63 28 58.425 &  185 & BS \\
1338+381 & 13 40 22.9519 & 37 54 43.839 &  105 & RS \\
1357+404 & 13 59 38.0943 & 40 11 38.260 &  129 & EF \\
1403+411 & 14 05 07.7949 & 40 56 57.847 &  134 & BS \\
1454+447 & 14 55 54.1361 & 44 31 37.668 &  115 & BS \\
1533+487 & 15 35 14.6540 & 48 36 59.697 &  122 & BS \\
1534+501 & 15 35 52.0395 & 49 57 39.084 &  103 & BS \\
1544+398 & 15 45 53.2331 & 39 41 46.857 &  168 &  G \\
1607+563 & 16 08 20.7518 & 56 13 56.373 &  129 & BS \\
1614+466 & 16 16 03.7667 & 46 32 25.231 &   98 & BS \\
1722+611 & 17 22 40.0578 & 61 05 59.801 &  213 & BS \\
1724+609 & 17 24 41.4142 & 60 55 55.731 &  179 & EF \\
1753+648 & 17 54 07.5904 & 64 52 02.642 &   89 & BS \\
1801+459 & 18 02 25.1427 & 45 57 34.645 &  192 & BS \\
1812+560 & 18 12 57.6692 & 56 03 49.198 &  193 & BS \\
1820+397 & 18 21 59.6991 & 39 45 59.647 &  743 & BS \\
1822+682 & 18 21 59.4951 & 68 18 43.003 &  192 & BS \\
1828+399 & 18 29 56.5203 & 39 57 34.690 &  116 & EF \\
1839+389 & 18 40 57.1550 & 39 00 45.712 &  133 & BS \\
1839+548 & 18 40 57.3780 & 54 52 15.920 &  121 & BS \\
1937+630 & 19 38 16.1680 & 63 07 17.803 &  254 &  ? \\
1939+429 & 19 40 49.3198 & 43 04 24.671 &  197 & BS \\
1941+413 & 19 42 58.6385 & 41 29 23.073 &  167 & BS \\
2000+472 & 20 02 10.4183 & 47 25 28.777 &  150 & CW \\
2005+642 & 20 06 17.6949 & 64 24 45.423 &  137 & RS \\
2013+508 & 20 14 28.5899 & 50 59 09.532 &  174 & EF \\
2014+463 & 20 15 39.9865 & 46 28 50.886 &  122 & CW \\
2112+374 & 21 14 44.1230 & 37 42 25.719 &  181 & CW \\
2119+709 & 21 19 54.1676 & 71 10 36.091 &  147 & EF \\
2151+431 & 21 53 50.9585 & 43 22 54.497 &  110 & NS \\
2202+716 & 22 03 30.4694 & 71 51 08.527 &  200 & RS \\
2248+555 & 22 50 42.8496 & 55 50 14.608 &  212 &  G \\
2300+638 & 23 02 41.3165 & 64 05 52.858 &  117 & EF \\
2310+724 & 23 12 19.6998 & 72 41 26.924 &  231 & EF \\
2341+697 & 23 43 43.7360 & 70 03 19.398 &  160 & NS \\
\noalign{\smallskip}
\hline
\end{tabular}
\end{flushleft}
\small{
Explanation of ID field is given in Table~1.\\
}
\end{table}
\begin{table}
\caption[ ]{Subsample Two --- sources not measured at 408~MHz}
\begin{flushleft}
\begin{tabular}{rrrc}
\hline\noalign{\smallskip}
IAUname & \multicolumn{1}{c}{R.A.} & \multicolumn{1}{c}{Dec.} & Opt. \\
\multicolumn{1}{c}{(B1950)} & \multicolumn{2}{c}{(J2000)} & ID\\
\noalign{\smallskip}
\hline\noalign{\smallskip}

0310+435 & 03 14 08.0539 & 43 45 19.770 & EF \\
0314+696 & 03 19 22.0734 & 69 49 25.603 & EF \\
0418+437 & 04 21 52.0619 & 43 53 04.216 & CW \\
0537+392 & 05 40 44.4377 & 39 16 12.236 & RS \\
0601+578 & 06 05 42.2275 & 57 53 16.351 & BS \\
0638+528 & 06 42 27.8215 & 52 47 59.282 & BS \\
0644+491 & 06 48 47.1190 & 49 07 20.736 & BS \\
0651+428 & 06 54 43.5263 & 42 47 58.728 &  G \\
0903+669 & 09 07 23.5240 & 66 44 46.942 & EF \\
1238+702 & 12 40 34.6989 & 69 58 30.616 & BS \\
1245+716 & 12 47 09.3270 & 71 24 20.018 & EF \\
1341+691 & 13 43 00.5520 & 68 55 17.160 & BS \\
1406+564 & 14 08 12.9466 & 56 13 32.488 & BS \\
1436+445 & 14 38 28.5048 & 44 18 12.085 & BS \\
1447+536 & 14 48 59.1739 & 53 26 09.282 & EF \\
1456+375 & 14 58 44.7949 & 37 20 21.627 & BS \\
1526+670 & 15 26 42.8732 & 66 50 54.617 & NS \\
1550+582 & 15 51 58.2077 & 58 06 44.466 & BS \\
1611+425 & 16 13 04.8038 & 42 23 18.903 & BS \\
1622+665 & 16 23 04.5221 & 66 24 01.084 &  G \\
1924+420 & 19 26 31.0504 & 42 09 58.991 &  G \\
2119+664 & 21 20 46.2045 & 66 42 20.216 & EF \\
2132+406 & 21 34 24.1053 & 40 50 11.345 & EF \\
2230+625 & 22 32 22.8655 & 62 49 36.436 & OB \\
2236+678 & 22 38 15.0284 & 68 04 59.758 & OB \\
2249+402 & 22 51 59.7715 & 40 30 58.155 & BS \\
2351+550 & 23 53 42.3011 & 55 18 40.670 & BS \\
\noalign{\smallskip}
\hline
\end{tabular}
\end{flushleft}
\small{
Explanation of ID field is given in Table~1.\\
}
\end{table}

\section{The new sample of GPS spectrum sources}

Finally we merged Subsample One (24~sources) with those 98~sources from
Subsample Two
whose flux densities we had successfully measured with MERLIN at 408~MHz. 
Since the selection process for this project was carried out, the 5~GHz
Green Bank survey of WB92 has been superseded by the GB6 survey (Gregory
et al.~1996) and at 1.4 GHz we can now use the the NRAO VLA Sky Survey
(NVSS)\footnote{available at ftp://nvss.cv.nrao.edu/pub/nvss/CATALOG}
(Condon et al.~1998) which has a resolution of $45\arcsec$ and an rms
sensitivity of approximately 1.5~mJy. 

Unfortunately, at the time of writing, NVSS --- although almost complete ---
did not cover all areas of the sky within its declination limits. Many NVSS
maps ($4\degr \times 4\degr$ each) appear to be "patchy" and the "holes" can
sometimes be quite large. Our survey suffered considerably from this
shortcoming of the current edition of NVSS --- 9~sources out of those
122~sources we wanted to study were simply not present in the NVSS catalogue.
(One source out of these nine was also unavailable in GB6.)
Additionally we decided to remove 2~other sources from the further
processing: one of these is blended with a nearby source and the second one
has an extended structure which should be studied in more detail.

At 1.4~GHz we
also tried to use the Faint Images of Radio Sky at Twenty (FIRST)
catalogue\footnote{available at http://sundog.stsci.edu/} (White et
al.~1997) --- 24~our sources could be found there. For 20~objects out of
these we noted a very good compatibility between FIRST and NVSS based fluxes;
the 4~objects which showed discrepancy are indicated in Table~6.

The selection process described above gave us finally 111~objects for which
we arrayed the flux density values at each frequency. Then we attempted to
fit model spectra to the available data using a broken power-law with 
the following formula (Moffet~1975):
 
\begin{equation}
S(\nu)={S_0\over1-e^{-1}} \cdot (\nu/\nu_0)^k \cdot 
(1-e^{-(\nu/\nu_0)^{l-k}}).
\end{equation}
 
Here $k$ and $l$ are the spectral indices of the rising and declining
parts of the spectrum as often used in radio astronomy, while $S_0$ and
$\nu_0$ are just fitting parameters which are {\it not} equal to the
maximum flux density ($S_{\max}$) and the peak frequency ($\nu_{\max}$) of
the fitted spectrum. Even though the broken power-law seems to be the
physically more sensible choice for a model spectrum compared to a
simple second-order polynomial, it has the disadvantage that it is
unconstrained if the peak of the spectrum falls beyond the 2nd highest
or below the 2nd lowest available frequency. In these cases we fixed
the peak of the model spectrum (i.e. $S_{\max}$ and $\nu_{\max}$) at the
peak of the measured data --- this was usually the measurement at
4.85~GHz --- and marked the fit as unconstrained in Table~4. This
means that the values for $k$ or $l$ have to be considered as a lower or
upper limit respectively (i.e. in reality the spectrum will be more
inverted at low frequencies or steeper at high frequencies).

The spectra of 35~sources could not be fitted with such a
convex-shaped curve (Fig~1.) and we claim that those sources cannot be
termed "GPS sources" at all and most likely are just variable
flat-spectrum sources. The 76~spectra that could be fitted with our
algorithm are presented in Fig.~2 and the fitting parameters are
given in Table~4. As can be seen from Fig.~2 and Table~4, some 
of the sources with unconstrained model spectra, fit the data relatively
poorly at low frequencies or have relatively flat spectral indices
(e.g.~0307+380 and 0610+510) and thus are less probable GPS candidates.

In Table~5 we specified some parameters of our "new" GPS sources gathered
from the literature: the names of other catalogues a particular source is
a member, the optical identification according to the NASA/IPAC
Extragalactic Database (NED) and the redshift. At the time of writing
21~objects from our collection have been identified (3~galaxies\footnote{
It is worth noting that our identifications shown in Tables 1 and 2
yielded 4~galaxies. The identifications for 0651+410 and 1245+676 are
given by NED, for the two other ones --- 1544+398 and 2248+555 --- are not.},
18~QSOs) and their redshifts are known. Those 3~galaxies have low
redshifts ($z\la0.1$); on the other hand --- as expected --- the majority
of quasars have large redshifts: for 6~QSOs $1<z\la 2$, for 7~other QSOs
$z\ga 2$. One QSO, 1338+381, is extremely redshifted: $z=3.103$. 

Most of the GPS sources studied so far hardly show any variability,
therefore we checked our sources against possible flux variations.
Firstly, because a significant variability of the flux density would mean
that the source in question is likely not to be a GPS and secondly ---
since our data are not simultaneous --- any variability makes derivation
of spectra questionable. The part of sources' spectra around 1.4~GHz is
obviously the most "sensitive" with regard to the GPS phenomenon so we
compared fluxes at this frequency given in WB92 to those from NVSS. We
applied corrections for the different beam sizes of these two measurements.
If a particular source had changed its flux between epochs of the GB
surveys and NVSS/FIRST more than 25~\% or the 1.4~GHz GB flux was missing
in WB92 we treated such a source as potentially variable, unless we could
find a second epoch flux density measurement elsewhere. We assigned a "candidate"
status for such objects and listed them in Table~6. Among these there are
4~sources (0412+447, 1125+366, 1357+404, 2005+642) with inverted spectra
only, i.e. apparently having turnovers in their spectra at frequencies
larger than 8.4~GHz. This feature was yet another reason to assign them a
candidate status. 

\section{Notes on individual sources}

Apart from the information in Table~5 we note that:

\begin{itemize}
\item 0627+532, 0652+577, 1107+485, 1256+546 and 1311+552 are 6C sources
(Hales et al.~1993 and references therein)
\item 1107+485 is a member of the DRAO 408~MHz survey (Green \& Riley~1995)
\item 1745+670 and 1753+648 are members of the NEP survey (Kollgaard et
al.~1994)
\item 1607+563, 1755+578 and 1815+614 are members of the 7C survey (Visser et
al.~1995)
\item 0514+474 was observed by Leahy \& Roger~(1996)
\item 0102+480 and 2253+417 are members of the CJ1 survey (Polatidis et
al.~1995)
\item 1245+676 is a giant radio galaxy with a GPS core (O'Dea priv. comm.~1996).
\item 0140+490 has a large scale symmetric structure $2\farcm5$ across.
\item 1839+548 and 2119+709 are marked as "quasi-point" sources in JVAS.
All other sources are pointlike i.e. they are unresolved by the VLA in
"A"~configuration at 8.4~GHz.
\item 1815+614 and 1946+708 are CSOs (Taylor et al.~1996, Taylor \&
Vermeulen~1997)
\end{itemize}

\section{Summary}

Gigahertz Peaked Spectrum objects are an astrophysically significant and
important class yet they are still not well understood (O'Dea~1998).
They are not necessarily a uniform class, and 
one can easily name subclasses among the whole GPS ensemble. For
example CSOs make one well defined group --- all of them are GPS galaxies
with characteristic VLBI morphologies. It is claimed by Readhead et
al.~(1996b) that CSOs play a key role as an initial stage in the
evolutionary scenario of radio-loud AGNs. 

Another fascinating subset of GPS class are objects with extreme ($z>3$)
redshifts. There are 9~such objects in the "working sample" (O'Dea, priv.
comm.~1996). Additionally there are $\sim 20$ objects with high ($1<z<3$)
redshifts. All the objects with $z>1$ are identified with quasars. 

The above two issues alone are already a good argument to extend the
number of known GPS radio sources. With such a goal in mind we have made a
search for candidate GPS sources in the part of the Jodrell Bank--VLA
Astrometric Survey limited to $35\degr \leq \delta \leq 75\degr$, namely
we compared 8.4~GHz flux densities derived from JVAS with respective
1.4~GHz and 5~GHz fluxes in available cata\-logues. We treated a source as
a plausible candidate if its spectrum seemed to be convex according to the
criterion we had arbitrarily assumed. 

Quite expectedly some of the candidates selected in this manner are
already recognised as GPS objects so we did not deal with them here. Using
flux densities at low frequencies i.e. 365~MHz and sometimes even 151~MHz
in available catalogues we were able to classify 24~objects as GPS sources
without any further measurements.  For the majority of selected objects
the flux densities at frequencies well below 1~GHz were not available.
In these cases we performed observations with MERLIN at 408~MHz
to establish the low-frequency part of the spectra. 

Our final decision on which of our candidates are and which are not GPS
sources has been made based on our MERLIN data (or Texas catalogue when
available), NVSS, GB6 and JVAS catalogues. Combining flux density
measurements made with high resolution both at low (MERLIN, 408~MHz) and
high (VLA, 8.4~GHz) frequencies enabled us to eliminate the effects of
confusion and any contribution from possible extended "halos".  We
regarded a source as a GPS if it had fitted well a "broken power-law"
function and was not variable. 

The sample we present here is the largest single contribution to the pool
of known GPS sources collected so far. Only 3~of our sources (0513+714,
0758+594, 1946+708) are overlapping with another large collection of GPS
sources, namely with the WENSS based sample\footnote{0513+714 and 0758+594
are not marked in Table 5 as WENSS sources because these objects are not
mentioned by Rengelink et al. (1997). On the other hand, except 1946+708,
the WENSS objects we list in Table 5 have {\it not} been recognised as GPS
by SSB98.} (SSB98). Seven sources in our sample have large redshifts
($z\ga 2$); the largest one is $z=3.103$. 

Our approach to finding GPS sources was to search from high frequency to
lower ones. WENSS has been equally successful in defining GPS sources but
searching from low frequencies to higher ones. We want to stress though
that our approach can successfully be applied to the areas of the sky {\it
not} planned to be covered by WENSS ($\delta < 30\degr$) but already
covered by the other catalogues we used (NVSS, GB6, JVAS). For the part of
JVAS we used so far (Patnaik et al.~1992) we found that 9\% of JVAS
sources are GPS; therefore the whole JVAS encompassing around 3000~sources
could easily yield 250 -- 300~of such objects. 

\begin{acknowledgements}

The initial stages of the programme described in this paper were completed
when AM stayed at MPIfR in Bonn. The Max-Planck-Gesellschaft stipend
which supported him in that time is gratefully acknowledged.

AM acknowledges support from the Polish State Committee for Scientific Research
grant 2.P304.003.07 and EU grant ERBCIPDCT940087.

HF is supported in part by the DFG, grant Fa 358/1-1\&2.

We thank the NVSS team led by Jim Condon for making NVSS public domain
data prior its final publication. Special thanks to Bill Cotton for his
co-operation. 

This research has made use of the NASA/IPAC Extragalactic Database (NED)
which is operated by the Jet Propulsion Laboratory, California Institute
of Technology, under contract with the National Aeronautics and Space
Administration. 

\end{acknowledgements}

\begin{figure*}

\epsfxsize=19cm
\epsfysize=23cm
\epsffile{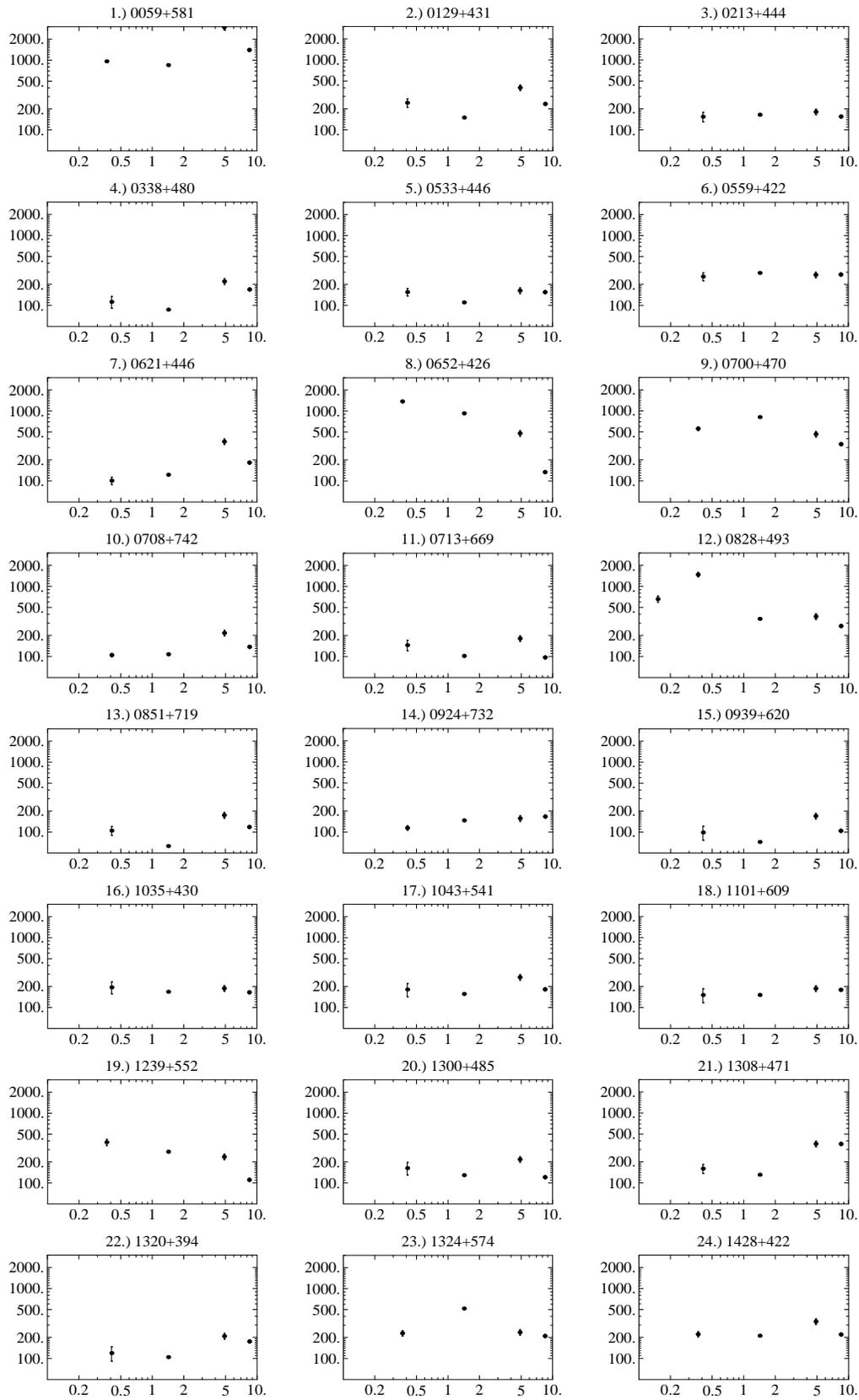}

\caption[]{Spectra of non-GPS sources. Abscissae: frequency in GHz, ordinates:
flux densities in mJy}
\end{figure*}
\begin{figure*}
 
\epsfxsize=19cm
\epsfysize=23cm
\epsffile{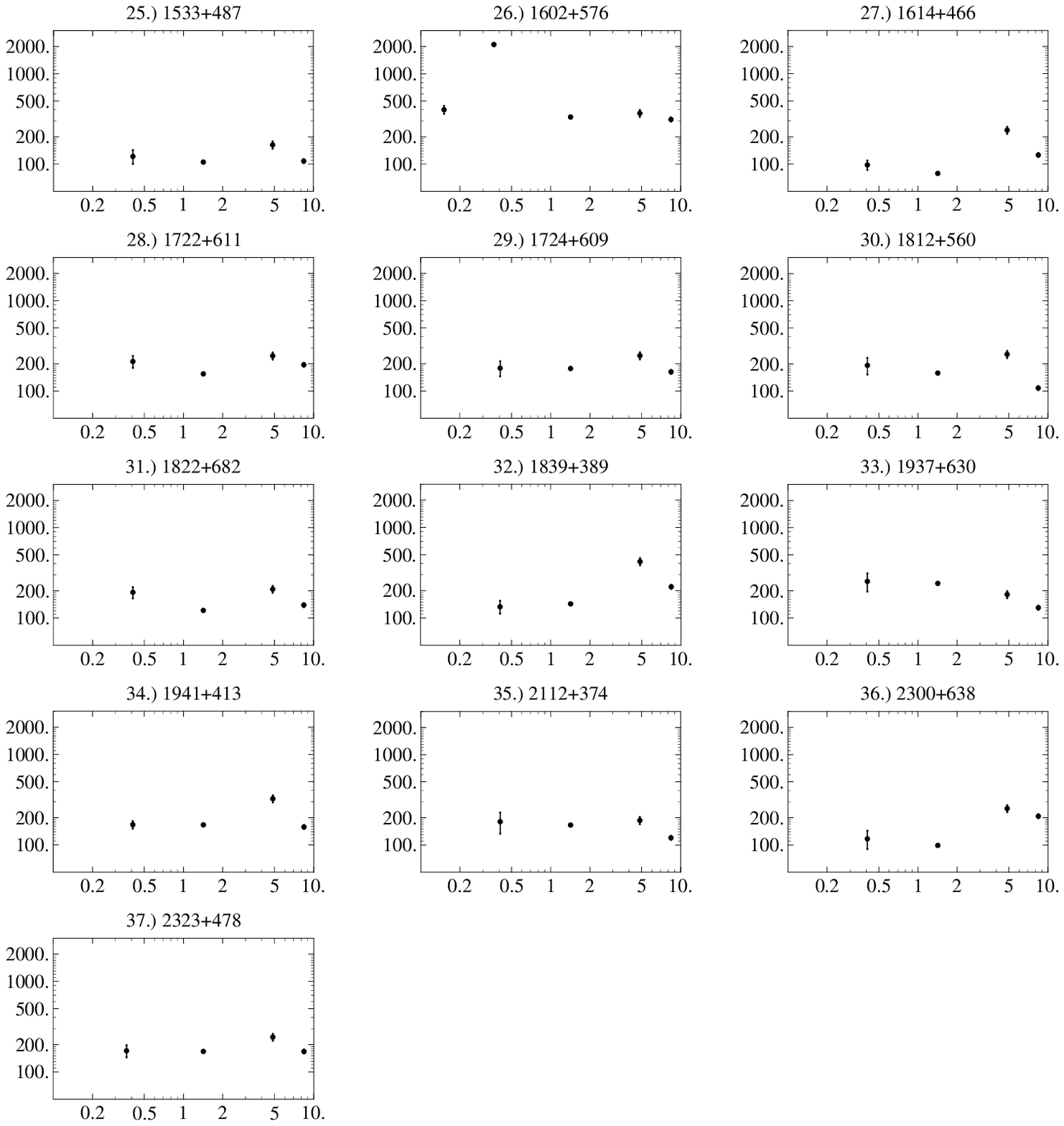}
 
{\small{\bf Fig. \thefigure{}.} continued}
\end{figure*}
\begin{figure*}

\epsfxsize=19cm
\epsfysize=23cm
\epsffile{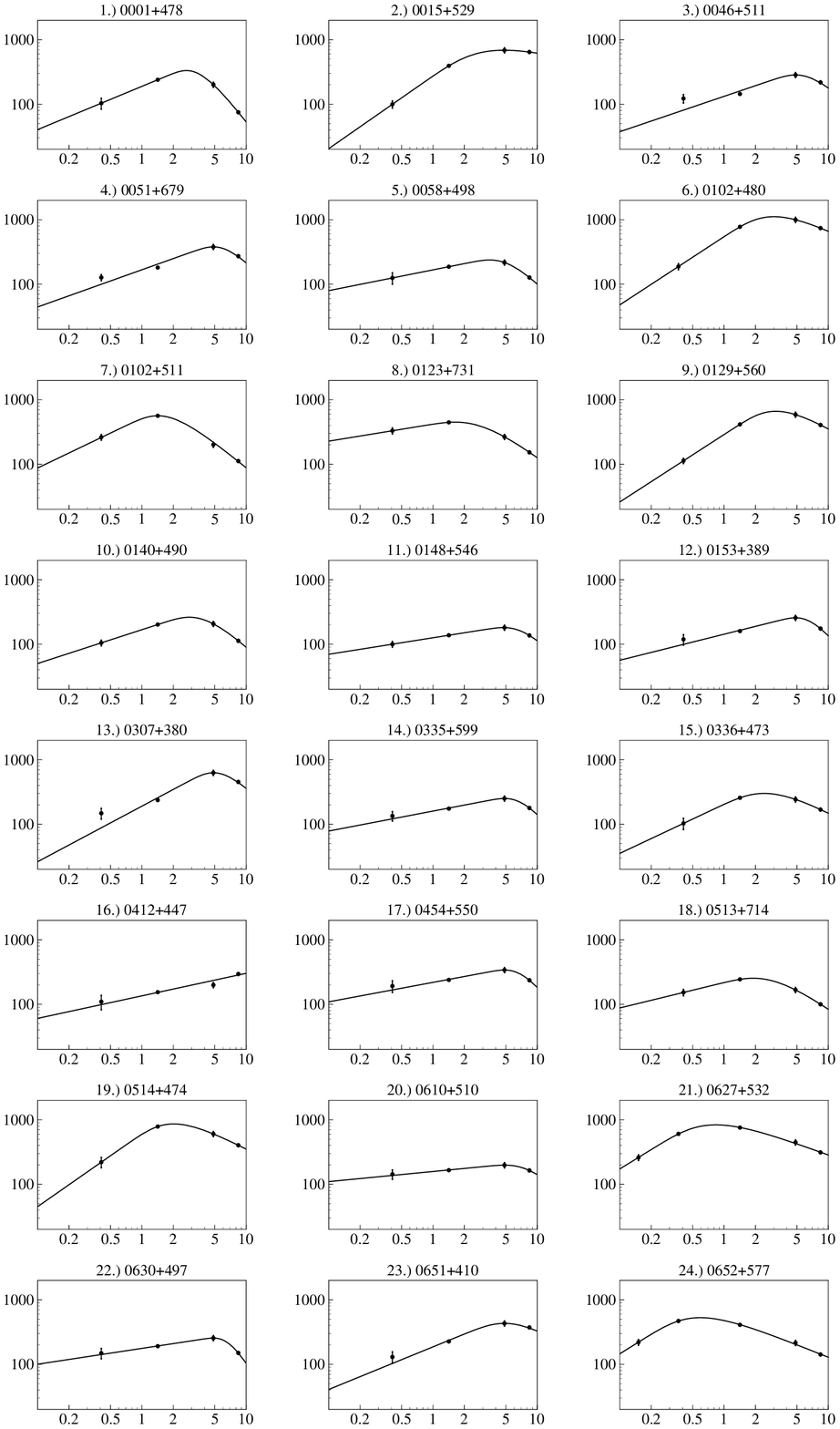}

\caption[]{Spectra of GPS sources. Abscissae: frequency in GHz, ordinates:
flux densities in mJy}
\end{figure*}
\begin{figure*}

\epsfxsize=19cm
\epsfysize=23cm
\epsffile{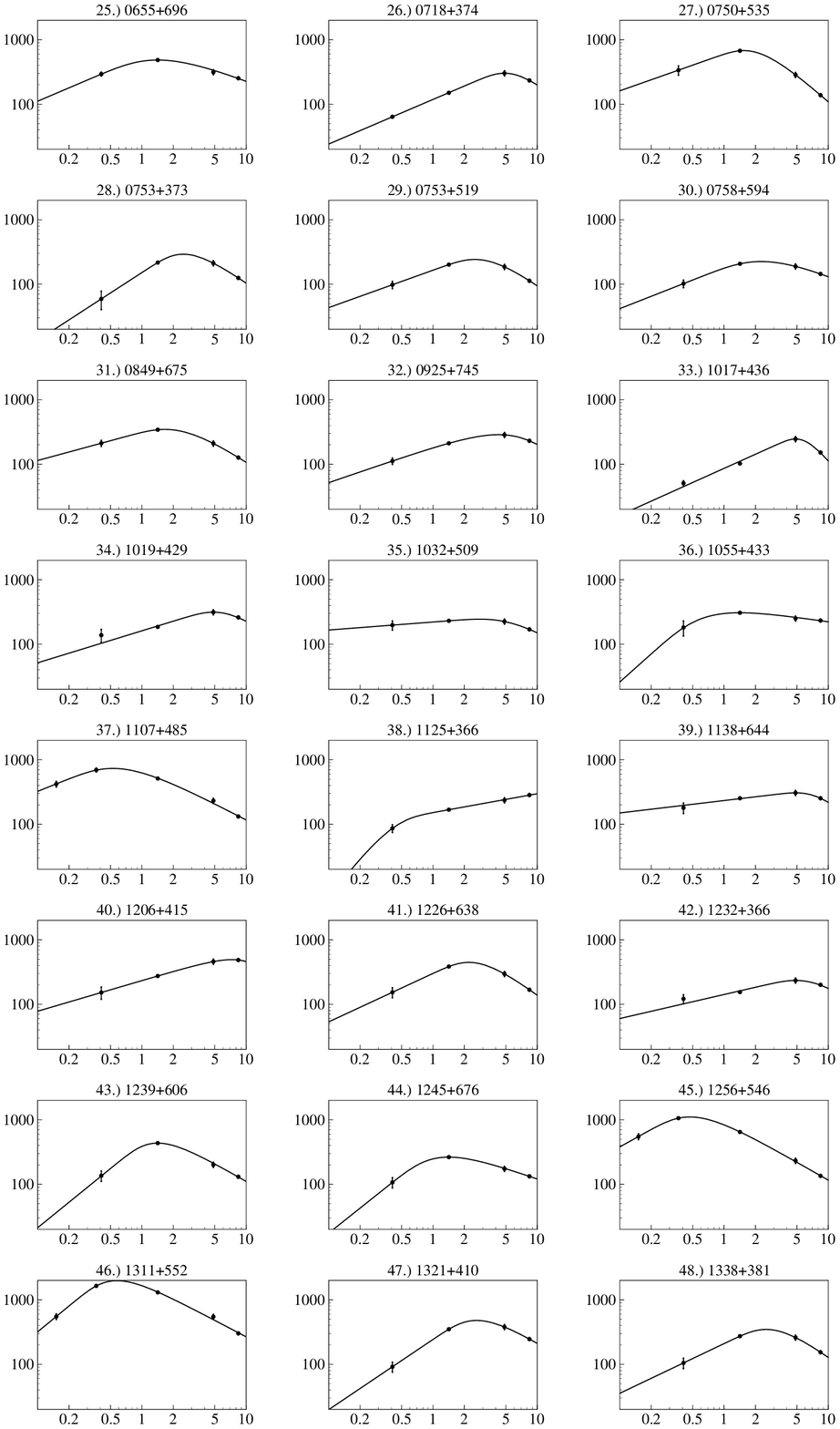}

{\small{\bf Fig. \thefigure{}.} continued}
\end{figure*}
\begin{figure*}

\epsfxsize=19cm
\epsfysize=23cm
\epsffile{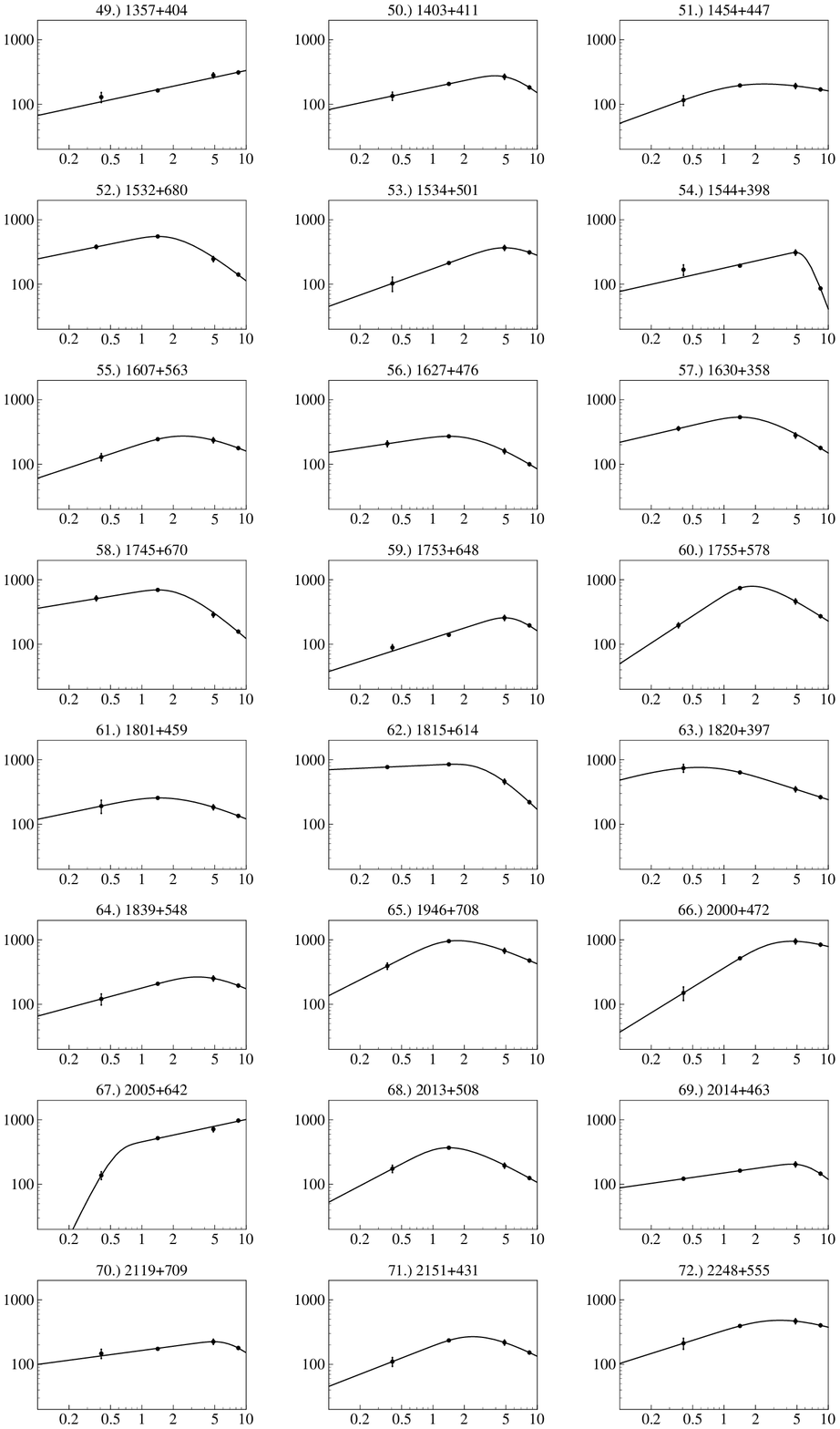}

{\small{\bf Fig. \thefigure{}.} continued}
\end{figure*}
\begin{figure*}

\epsfxsize=19cm
\epsfysize=23cm
\epsffile{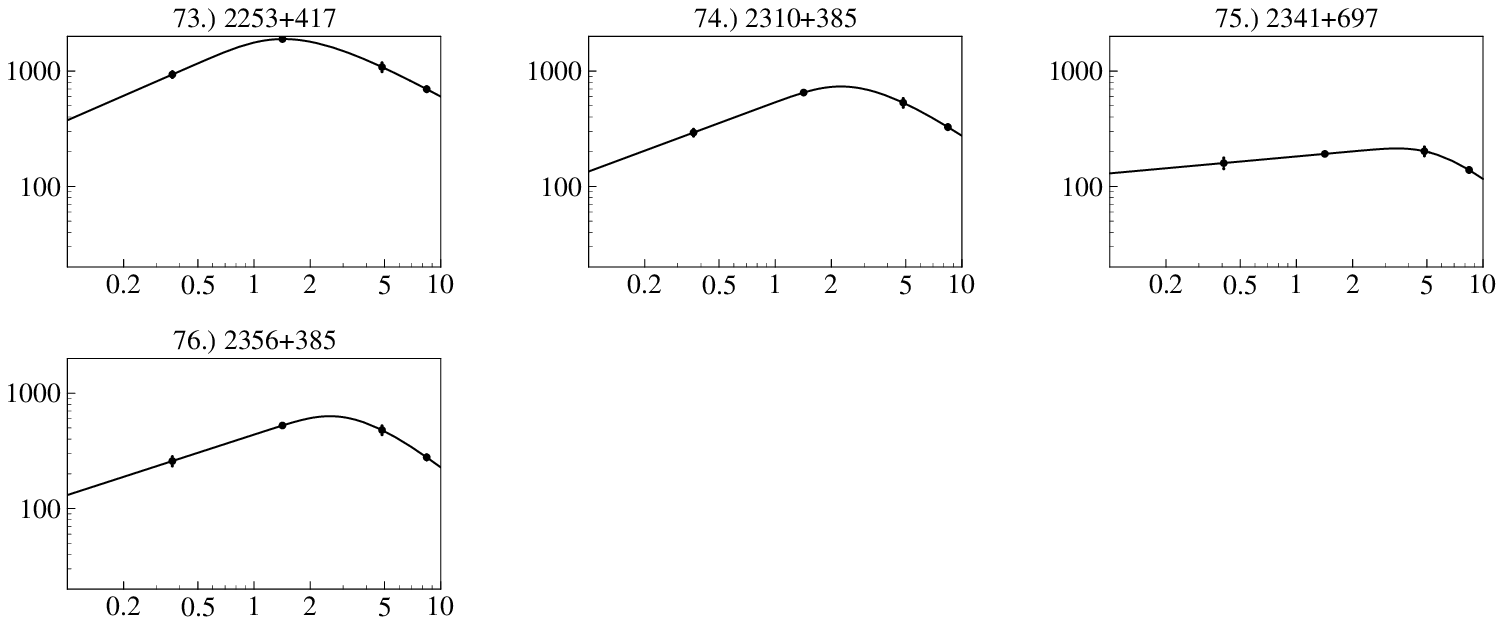}

{\small{\bf Fig. \thefigure{}.} continued}
\end{figure*}

\begin{table*}
\begin{flushleft}
\caption[ ]{GPS sources' spectra fitting parameters}
\begin{tabular}{rlrlllrrc}
\hline\noalign{\smallskip}
Number & B1950name & \multicolumn{1}{c}{$S_0$} & \multicolumn{1}{c}{$\nu_0$} &
\multicolumn{1}{c}{$k$} & \multicolumn{1}{c}{$l$} & $S_{max}$ & $\nu_{max}$ 
&  unconstrained \\ \noalign{\smallskip}
\hline\noalign{\smallskip}

1  & 0001+478 & 286.  & 3.66  & $+$0.671  & $-$2.1   & 332.  & 2.68 &     \\
2  & 0015+529 & 657.  & 3.27  & $+$1.13   & $-$0.381 & 687.  & 4.82 &     \\
3  & 0046+511 & 246.  & 7.14  & $+$0.549  & $-$1.63  & 284.  & 4.85 &  $\bullet$ 
 \\
4  & 0051+679 & 324.  & 6.97  & $+$0.577  & $-$1.85  & 379.  & 4.85 &  $\bullet$ 
 \\
5  & 0058+498 & 187.  & 5.94  & $+$0.322  & $-$1.76  & 237.  & 3.49 &     \\
6  & 0102+480 & 1110. & 3.03  & $+$1.06   & $-$0.774 & 1110. & 3.01 &     \\
7  & 0102+511 & 530.  & 1.85  & $+$0.775  & $-$1.33  & 564.  & 1.42 &  $\bullet$ 
 \\
8  & 0123+731 & 361.  & 3.25  & $+$0.264  & $-$1.27  & 449.  & 1.63 &     \\
9  & 0129+560 & 659.  & 3.42  & $+$1.05   & $-$0.956 & 662.  & 3.14 &     \\
10 & 0140+490 & 228.  & 4.28  & $+$0.524  & $-$1.53  & 262.  & 2.85 &     \\
11 & 0148+546 & 137.  & 8.39  & $+$0.256  & $-$1.86  & 181.  & 4.73 &     \\
12 & 0153+389 & 203.  & 7.44  & $+$0.404  & $-$2.17  & 257.  & 4.85 &  $\bullet$ 
 \\
13 & 0307+380 & 581.  & 6.21  & $+$0.862  & $-$1.66  & 628.  & 4.85 &  $\bullet$ 
 \\
14 & 0335+599 & 192.  & 7.92  & $+$0.309  & $-$2.12  & 251.  & 4.85 &  $\bullet$ 
 \\
15 & 0336+473 & 295.  & 2.96  & $+$0.762  & $-$0.886 & 301.  & 2.43 &     \\
16 & 0412+447 & 277.  & 7.84  & $+$0.349  & $+$0.349  &       &      &     \\
17 & 0454+550 & 257.  & 7.85  & $+$0.302  & $-$2.23  & 340.  & 4.85 &  $\bullet$ 
 \\
18 & 0513+714 & 219.  & 3.26  & $+$0.394  & $-$1.19  & 253.  & 1.91 &     \\
19 & 0514+474 & 863.  & 2.05  & $+$1.13   & $-$0.842 & 863.  & 2.02 &     \\
20 & 0610+510 & 143.  & 9.9   & $+$0.157  & $-$1.74  & 198.  & 4.85 &  $\bullet$ 
 \\
21 & 0627+532 & 832.  & 0.767 & $+$0.996  & $-$0.593 & 835.  & 0.836&     \\
22 & 0630+497 & 183.  & 7.51  & $+$0.245  & $-$2.89  & 255.  & 4.85 &  $\bullet$ 
 \\
23 & 0651+410 & 408.  & 6.57  & $+$0.661  & $-$1.06  & 431.  & 4.85 &  $\bullet$ 
 \\
24 & 0652+577 & 528.  & 0.564 & $+$1.01   & $-$0.649 & 528.  & 0.592&     \\
25 & 0655+696 & 481.  & 1.63  & $+$0.69   & $-$0.644 & 484.  & 1.42 &  $\bullet$ 
 \\
26 & 0718+374 & 277.  & 6.62  & $+$0.692  & $-$1.43  & 302.  & 4.85 &  $\bullet$ 
 \\
27 & 0750+535 & 602.  & 2.26  & $+$0.57   & $-$1.44  & 678.  & 1.55 &     \\
28 & 0753+373 & 287.  & 2.85  & $+$1.05   & $-$1.15  & 292.  & 2.5  &     \\
29 & 0753+519 & 221.  & 3.66  & $+$0.58   & $-$1.22  & 242.  & 2.52 &     \\
30 & 0758+594 & 221.  & 2.84  & $+$0.635  & $-$0.711 & 225.  & 2.27 &     \\
31 & 0849+675 & 308.  & 2.72  & $+$0.44   & $-$1.11  & 347.  & 1.66 &     \\
32 & 0925+745 & 284.  & 4.95  & $-$1.07   & $+$0.554  & 287.  & 4.26 &     \\
33 & 1017+436 & 212.  & 6.49  & $+$0.728  & $-$2.22  & 246.  & 4.85 &  $\bullet$ 
 \\
34 & 1019+429 & 277.  & 7.5   & $+$0.496  & $-$1.3   & 314.  & 4.85 &  $\bullet$ 
 \\
35 & 1032+509 & 180.  & 7.71  & $+$0.124  & $-$1.16  & 245.  & 2.81 &     \\
36 & 1055+433 & 267.  & 0.69  & $+$1.44   & $-$0.237 & 308.  & 1.42 &  $\bullet$ 
 \\
37 & 1107+485 & 712.  & 0.69  & $+$0.645  & $-$0.845 & 734.  & 0.528&     \\
38 & 1125+366 & 70.4  & 0.347 & $+$0.292  & $+$2.14   &       &      &     \\
39 & 1138+644 & 228.  & 9.59  & $+$0.192  & $-$1.68  & 307.  & 4.85 &  $\bullet$ 
 \\
40 & 1206+415 & 449.  & 10.4  & $-$2.04   & $+$0.475  & 492.  & 7.29 &     \\
41 & 1226+638 & 422.  & 2.89  & $+$0.751  & $-$1.24  & 446.  & 2.21 &     \\
42 & 1232+366 & 198.  & 8.41  & $+$0.372  & $-$1.29  & 234.  & 4.85 &  $\bullet$ 
 \\
43 & 1239+606 & 435.  & 1.41  & $+$1.31   & $-$0.93  & 435.  & 1.42 &  $\bullet$ 
 \\
44 & 1245+676 & 258.  & 1.16  & $+$1.29   & $-$0.561 & 264.  & 1.42 &  $\bullet$ 
 \\
45 & 1256+546 & 1090. & 0.539 & $+$0.901  & $-$0.925 & 1110. & 0.471&     \\
46 & 1311+552 & 1980. & 0.555 & $+$1.33   & $-$0.851 & 1980. & 0.578&     \\
47 & 1321+410 & 476.  & 2.83  & $+$1.09   & $-$0.98  & 478.  & 2.62 &     \\
48 & 1338+381 & 330.  & 3.27  & $+$0.771  & $-$1.21  & 347.  & 2.53 &     \\
49 & 1357+404 & 356.  & 12.1  & $+$0.349  & $+$0.349  &       &      &     \\
50 & 1403+411 & 223.  & 6.71  & $+$0.346  & $-$1.58  & 276.  & 3.92 &     \\
\noalign{\smallskip}
\hline
\end{tabular}
\end{flushleft}
\end{table*}
\begin{table*}
{\small{\bf Table \thetable{}.} continued}
\begin{flushleft}
\begin{tabular}{rlrlllrrc}
\hline\noalign{\smallskip}
Number & B1950name & \multicolumn{1}{c}{$S_0$} & \multicolumn{1}{c}{$\nu_0$} &
\multicolumn{1}{c}{$k$} & \multicolumn{1}{c}{$l$} & $S_{max}$ & $\nu_{max}$
&  unconstrained \\ \noalign{\smallskip}
\hline\noalign{\smallskip}

51 & 1454+447 & 206.  & 2.36  & $+$0.588  & $-$0.407 & 206.  & 2.43 &     \\
52 & 1532+680 & 455.  & 2.57  & $+$0.329  & $-$1.32  & 551.  & 1.42 &  $\bullet$ 
 \\
53 & 1534+501 & 341.  & 6.97  & $+$0.584  & $-$1.09  & 367.  & 4.85 &  $\bullet$ 
 \\
54 & 1544+398 & 221.  & 6.4   & $+$0.363  & $-$4.67  & 310.  & 4.85 &  $\bullet$ 
 \\
55 & 1607+563 & 261.  & 3.54  & $+$0.538  & $-$0.797 & 273.  & 2.48 &     \\
56 & 1627+476 & 219.  & 3.06  & $+$0.241  & $-$1.11  & 271.  & 1.42 &  $\bullet$ 
 \\
57 & 1630+358 & 466.  & 2.5   & $+$0.377  & $-$1.11  & 537.  & 1.42 &  $\bullet$ 
 \\
58 & 1745+670 & 548.  & 2.7   & $+$0.266  & $-$1.46  & 696.  & 1.42 &  $\bullet$ 
 \\
59 & 1753+648 & 220.  & 7.26  & $+$0.519  & $-$1.66  & 257.  & 4.85 &  $\bullet$ 
 \\
60 & 1755+578 & 785.  & 2.07  & $+$1.06   & $-$1.07  & 793.  & 1.86 &     \\
61 & 1801+459 & 231.  & 2.78  & $+$0.337  & $-$0.77  & 256.  & 1.46 \\
62 & 1815+614 & 580.  & 3.9   & $+$0.0731 & $-$1.69  & 852.  & 1.61 &     \\
63 & 1820+397 & 753.  & 0.457 & $-$0.519  & $+$0.527  & 761.  & 0.575&     \\
64 & 1839+548 & 240.  & 5.62  & $+$0.436  & $-$0.983 & 265.  & 3.42 &     \\
65 & 1946+708 & 965.  & 1.92  & $+$0.818  & $-$0.755 & 970.  & 1.72 &     \\
66 & 2000+472 & 946.  & 4.11  & $+$0.996  & $-$0.588 & 949.  & 4.51 &     \\
67 & 2005+642 & 227.  & 0.502 & $+$0.346  & $+$3.4    &       &      &     \\
68 & 2013+508 & 363.  & 1.67  & $+$0.849  & $-$0.927 & 369.  & 1.42 &  $\bullet$ 
 \\
69 & 2014+463 & 153.  & 8.03  & $+$0.231  & $-$1.88  & 205.  & 4.44 &     \\
70 & 2119+709 & 167.  & 9.07  & $+$0.217  & $-$1.79  & 224.  & 4.85 &  $\bullet$ 
 \\
71 & 2151+431 & 257.  & 3.25  & $+$0.63   & $-$0.917 & 268.  & 2.41 &     \\
72 & 2248+555 & 469.  & 4.72  & $+$0.512  & $-$0.633 & 481.  & 3.43 &     \\
73 & 2253+417 & 1840. & 1.83  & $+$0.706  & $-$0.909 & 1900. & 1.44 &     \\
74 & 2310+385 & 682.  & 3.2   & $+$0.6    & $-$1.14  & 735.  & 2.25 &     \\
75 & 2341+697 & 155.  & 7.5   & $+$0.147  & $-$1.59  & 213.  & 3.45 &     \\
76 & 2356+385 & 560.  & 3.85  & $+$0.524  & $-$1.34  & 632.  & 2.55 &     \\
\noalign{\smallskip}
\hline
\end{tabular}
\end{flushleft}
\end{table*}

\begin{table*}
\begin{flushleft}
\caption[ ]{Some other parameters of JVAS GPS sources}
\begin{tabular}{rlccccccll}
\hline\noalign{\smallskip}
Number & B1950name & S4/S5 & FIRST & B3 & WENSS & CJ2 & Opt. ID & $z$ & Redshift
 reference\\
\noalign{\smallskip}
\hline\noalign{\smallskip}
 
1  & 0001+478 & & & & & & & & \\
2  & 0015+529 & & & & & & & & \\
3  & 0046+511 & & & & & & & & \\
4  & 0051+679 & & & & & & & & \\
5  & 0058+498 & & & & & & & & \\
6  & 0102+480 & $\bullet$ & & & & & & & \\
7  & 0102+511 & & & & & & & & \\
8  & 0123+731 & $\bullet$ & & & & & & & \\
9  & 0129+560 & & & & & & & & \\
10 & 0140+490 & & & & & & & & \\
11 & 0148+546 & & & & & & & & \\
12 & 0153+389 & & & & & & & & \\
13 & 0307+380 & & & $\bullet$ & & $\bullet$ & QSO & 0.816 & Vermeulen \& Taylor, 
1995 \\
14 & 0335+599 & & & & & & & & \\
15 & 0336+473 & & & & & & & & \\
16 & 0412+447 & & & & & & & & \\
17 & 0454+550 & & & & & & & & \\
18 & 0513+714 & & & & & & & & \\
19 & 0514+474 & & & $\bullet$ & & & & & \\
20 & 0610+510 & & & & & & QSO & 1.59 & Hook et al., 1996 \\
21 & 0627+532 & $\bullet$ & & & & $\bullet$ & QSO & 2.204 & Henstock et al., 
1997 \\
22 & 0630+497 & $\bullet$ & & & & & & & \\
23 & 0651+410 & & & & & $\bullet$ & G & 0.02156 & Marzke et al., 1996 \\
24 & 0652+577 & & & & & & & & \\
25 & 0655+696 & & & & & & QSO & 1.971 & Moran et al., 1996 \\
26 & 0718+374 & & $\bullet$ & & & & & & \\
27 & 0750+535 & & $\bullet$ & & & & & & \\
28 & 0753+373 & & $\bullet$ & & & & & & \\
29 & 0753+519 & & $\bullet$ & & & & QSO & 1.33 & Hook et al., 1996 \\
30 & 0758+594 & & & & & & & & \\
31 & 0849+675 & & & & & & & & \\
32 & 0925+745 & $\bullet$ & & & & & & & \\
33 & 1017+436 & & $\bullet$ & & & & QSO & 1.96 & Hook et al., 1996 \\
34 & 1019+429 & $\bullet$ & $\bullet$ & $\bullet$ & & & & & \\
35 & 1032+509 & & $\bullet$ & & & & & & \\
36 & 1055+433 & $\bullet$ & $\bullet$ & $\bullet$ & & & & & \\
37 & 1107+485 & $\bullet$ & $\bullet$ & & & & QSO & 0.74 & Hook et al., 1996 \\
38 & 1125+366 & & $\bullet$ & & & & & & \\
39 & 1138+644 & & & & & & & & \\
40 & 1206+415 & & $\bullet$ & & & $\bullet$ & & & \\
41 & 1226+638 & & & & & & & & \\
42 & 1232+366 & $\bullet$ & $\bullet$ & & & & QSO & 1.60 & Hook et al., 1996 \\
43 & 1239+606 & & & & & & & & \\
44 & 1245+676 & & & & & & G & 0.103 & Marzke et al., 1996 \\
45 & 1256+546 & & $\bullet$ & & & & & & \\
46 & 1311+552 & $\bullet$ & $\bullet$ & & & $\bullet$ & QSO & 0.613 & Vermeulen et al., 
1996 \\
47 & 1321+410 & $\bullet$ & $\bullet$ & & & $\bullet$ & QSO & 0.496 & Vermeulen 
et al., 1996 \\
48 & 1338+381 & $\bullet$ & $\bullet$ & & & & QSO & 3.103 & Hook et al., 1995 \\
49 & 1357+404 & & $\bullet$ & & & & & & \\
50 & 1403+411 & & $\bullet$ & $\bullet$ & & & & & \\
\noalign{\smallskip}
\hline
\end{tabular}
\end{flushleft}
\end{table*}
\begin{table*}
{\small{\bf Table \thetable{}.} continued}
\begin{flushleft}
\begin{tabular}{rlccccccll}
\hline\noalign{\smallskip}
Number & B1950name & S4/S5 & FIRST & B3 & WENSS & CJ2 & Opt. ID & $z$ & Redshift 
reference\\
\noalign{\smallskip}
\hline\noalign{\smallskip}

51 & 1454+447 & & $\bullet$ & & & & & & \\
52 & 1532+680 & & & & $\bullet$ & & & & \\
53 & 1534+501 & $\bullet$ & $\bullet$ & & & $\bullet$ & QSO & 1.119 & Vermeulen \& Taylor, 
1995 \\
54 & 1544+398 & $\bullet$ & $\bullet$ & $\bullet$ & & & & & \\
55 & 1607+563 & & $\bullet$ & & & & & & \\
56 & 1627+476 & & $\bullet$ & & & & & & \\
57 & 1630+358 & & $\bullet$ & & & & & & \\
58 & 1745+670 & & & & $\bullet$ & & & & \\
59 & 1753+648 & & & & $\bullet$ & & & & \\
60 & 1755+578 & $\bullet$ & & & $\bullet$ & $\bullet$ & QSO & 2.110 & Henstock 
et al., 1997 \\
61 & 1801+459 & & & $\bullet$ & & & & & \\
62 & 1815+614 & & & & $\bullet$ & $\bullet$ & QSO & 0.601 & Vermeulen \& Taylor, 
1995 \\
63 & 1820+397 & $\bullet$ & & $\bullet$ & & & & & \\
64 & 1839+548 & & & & & & & & \\
65 & 1946+708 & $\bullet$ & & & $\bullet$ & $\bullet$ & G & 0.101 & Stickel \& 
K\"uhr, 1993 \\
66 & 2000+472 & & & & & & & & \\
67 & 2005+642 & & & & $\bullet$ & $\bullet$ & QSO & 1.574 & Henstock et al., 1997 \\
68 & 2013+508 & & & & & & & & \\
69 & 2014+463 & & & & & & & & \\
70 & 2119+709 & & & & & & & & \\
71 & 2151+431 & & & & & & & & \\
72 & 2248+555 & & & & & & & & \\
73 & 2253+417 & $\bullet$ & & & & & QSO & 1.476 & Hewitt \& Burbidge, 1989 \\
74 & 2310+385 & & & & & $\bullet$ & QSO & 2.17 & Hewitt \& Burbidge, 1989 \\
75 & 2341+697 & & & & & & & & \\
76 & 2356+385 & $\bullet$ & & & & $\bullet$ & QSO & 2.704 & Stickel \& K\"uhr, 
1994 \\
\noalign{\smallskip}
\hline
\end{tabular}
\end{flushleft}
\small{References of the catalogues:\\
\\
S4 --- Pauliny-Toth et al. 1978\\
S5 --- K\"uhr et al. 1981b\\
FIRST --- White et al. 1997\\
B3 --- Ficarra et al. 1985\\
WENSS --- Rengelink et al. 1997\\
CJ2 --- Taylor et al. 1994\\
}
\end{table*}

\begin{table*}
\begin{flushleft}
\caption[ ]{Candidate GPS sources}
\begin{tabular}{ll}
\hline\noalign{\smallskip}
B1950name & Reason \\
\noalign{\smallskip}  
\hline\noalign{\smallskip}

0001+478 & GB flux is 73\% greater than NVSS flux \\
0046+511 & no 2nd epoch data \\
0051+679 & GB flux is 36\% greater than NVSS flux \\
0058+498 & no 2nd epoch data \\
0307+380 & poor fit to the broken power-law curve \\
0412+447 & GB flux is 4.7 times greater than NVSS flux, inverted spectrum only 
\\
0651+410 & no 2nd epoch data \\
0655+696 & GB flux is 55\% less than NVSS flux \\
0718+374 & no 2nd epoch data \\
0753+519 & no 2nd epoch data \\
0758+594 & GB flux is 34\% less than NVSS flux \\
1017+436 & no 2nd epoch data \\
1055+433 & B3 flux (420~mJy) and MERLIN 408~MHz flux (182~mJy) do not match \\
1125+366 & FIRST flux does not match NVSS and GB fluxes, inverted spectrum only 
\\
1206+415 & FIRST flux does not match NVSS and GB fluxes \\
1232+366 & FIRST flux matches NVSS flux but they both don't match GB and S4 
fluxes --- \\
 & extended component? \\
1357+404 & FIRST flux does not match NVSS and GB fluxes, inverted spectrum only 
\\
1544+398 & B3 flux matches MERLIN flux but FIRST flux does not match NVSS and GB 
fluxes \\
1839+548 & no 2nd epoch data \\
2005+642 & inverted spectrum only \\
2013+508 & GB flux is 48\% less than NVSS flux \\
2014+463 & no 2nd epoch data \\
2248+555 & no 2nd epoch data \\
2341+697 & no 2nd epoch data \\
\noalign{\smallskip}
\hline
\end{tabular}
\end{flushleft}
\end{table*}


\begin{thebibliography}{}

\bibitem[]{}Baars J.W.M., Genzel R., Pauliny-Toth I.I.K., Witzel A., 1977, 
        A\&A 61, 99
\bibitem[]{}Cersosimo J.C., Lebron Santos M., Cintr\'on S.I., Quiniento Z.M.,
	1994, ApJS 95, 157
\bibitem[]{}Condon J.J., Cotton W.D., Greisen E.W., et al., 1998, AJ 115, 1693
\bibitem[]{}Douglas J.N., Bash F.N., Arakel Bozyan F., Torrence G.W., 1996,
	AJ 111, 1945
\bibitem[]{}Ficarra A., Grueff G., Tomasetti G., 1985, A\&AS 59, 255
\bibitem[]{}Gopal-Krishna, Patnaik A.R., Steppe H., 1983, A\&A 123, 107
\bibitem[]{}Gopal-Krishna, Spoelstra T.A.Th., 1993, A\&A 271, 103
\bibitem[]{}Green D.A., Riley J.M., 1995, MNRAS 274, 324
\bibitem[]{}Gregory P.C., Scott W.K, Douglas K., Condon J.J., 1996,
	ApJS 103, 427
\bibitem[]{}Hales S.E.G., Baldwin J.E., Warner P.J., 1993, MNRAS 263, 25
\bibitem[]{}Henstock D.R., Browne I.W.A., Wilkinson P.N., McMahon R.G., 1997,
	MNRAS 290, 380
\bibitem[]{}Hewitt A., Burbidge G., 1989, magnetic tape
\bibitem[]{}Hook I.M., McMahon R.G., Patnaik A.R., et al., 1995, MNRAS 273, 63
\bibitem[]{}Hook I.M., McMahon R.G., Irwin M.J., Hazard C., 1996,
	MNRAS 282, 1274
\bibitem[]{}Kollgaard R.I., Brinkmann W., McMath Chester M., et al., 1994,
	ApJS 93, 145
\bibitem[]{}K\"uhr H., Witzel A., Pauliny-Toth I.I.K., Nauber U., 1981a,
	A\&AS 45, 367
\bibitem[]{}K\"uhr H., Pauliny-Toth I.I.K., Witzel A., Schmidt J., 1981b,
	AJ 86, 854
\bibitem[]{}Leahy D.A., Roger R.S., 1996, A\&AS 115, 345
\bibitem[]{}Marzke R.O., Huchra J.P., Geller M.J., 1996, AJ 112, 1803
\bibitem[]{}Moffet A.T., 1975, in: Sandage A., Sandage M., Kristian J. (eds.)
	Stars and Stellar Systems, Vol. IX, p.211
\bibitem[]{}Moran E.C., Helfand D.J., Becker R.H., White R.L., 1996,
	ApJ 461, 127 
\bibitem[]{}O'Dea C.P., Baum S.A., Stanghellini C., 1991, ApJ 380, 66 (OBS91)
\bibitem[]{}O'Dea C.P., 1998, PASP 110, 493
\bibitem[]{}Patnaik A.R., Browne I.W.A., Wilkinson P.N., Wrobel J.M., 1992,
	MNRAS 254, 655
\bibitem[]{}Pauliny-Toth I.I.K., Witzel A., Preuss E., et al., 1978, AJ 83, 451
\bibitem[]{}Polatidis A.G., Wilkinson P.N., Xu W., et al., 1995, ApJS 98, 1
\bibitem[]{}Rengelink R.B., Tang Y., de Bruyn A.G., et al., 1997, A\&A 124, 259
\bibitem[]{}Readhead A.C.S., Taylor G.B., Xu W., et al., 1996a, ApJ 460, 612
\bibitem[]{}Readhead A.C.S., Taylor G.B., Pearson T.J., Wilkinson P.N., 1996b,
	ApJ 460, 634
\bibitem[]{}Snellen I.A.G., Zhang M., Schilizzi R.T., et al., 1995,
	A\&A 300, 359
\bibitem[]{}Snellen I.A.G., Schilizzi R.T., de Bruyn A.G., et al., 1998, A\&AS 131, 435 (SSB98)
\bibitem[]{}Spoelstra T.A.Th., Patnaik A.R., Gopal-Krishna, 1985, A\&A 152, 38
\bibitem[]{}Stanghellini C., O'Dea C.P., Baum S.A., Fanti R., 1990,
	{\it Compact Steep Spectrum \& Gigahertz-Peaked Spectrum Sources},
	Dwingeloo Workshop, 18--19 June 1990,
	Fanti C., Fanti R., O'Dea C.P., Schilizzi R.T. (eds.)
\bibitem[]{}Stanghellini C., Dallacasa D., O'Dea C.P., et al., 1996,
	{\it 2nd workshop on GPS and CSS Radio Sources},
	Leiden, 30 Sept.--- 2 Oct. 1996, eds. Snellen I.A.G.,
	Schilizzi R.T., R\"ottgering H.J.A., Bremer M.N.
\bibitem[]{}Stickel M., K\"uhr H., 1993, A\&AS 100, 395
\bibitem[]{}Stickel M., K\"uhr H., 1994, A\&AS 105, 67
\bibitem[]{}Taylor G.B., Vermeulen R.C., 1997, ApJ 485, 9
\bibitem[]{}Taylor G.B., Vermeulen R.C., Pearson T.J., et al., 1994,
	ApJS 95, 345
\bibitem[]{}Taylor G.B., Vermeulen R.C., Readhead A.C.S., et al., 1996,
	{\it 2nd workshop on GPS and CSS Radio Sources},
	Leiden, 30 Sept.--- 2 Oct. 1996, eds. Snellen I.A.G.,
        Schilizzi R.T., R\"ottgering H.J.A., Bremer M.N.
\bibitem[]{}Vermeulen R.C., Taylor G.B., 1995, AJ 109, 1983
\bibitem[]{}Vermeulen R.C., Taylor G.B., Readhead A.C.S., Browne I.W.A., 1996,
	AJ 111, 1013
\bibitem[]{}Visser A.E., Riley J.M., R\"ottgering H.J.A., Waldram E.M., 1995,
	A\&AS 110, 419
\bibitem[]{}de Vries W.H., Barthel P.D., O'Dea C.P., 1997, A\&A 321, 105
\bibitem[]{}Walker R.C., 1981, AJ 86, 1323
\bibitem[]{}Wilkinson P.N., Polatidis A.G., Readhead A.C.S., Xu W., Pearson 
T.J.,
	1994, ApJ 432, L87
\bibitem[]{}White R.L., Becker R. H., 1992, ApJS 79, 331 (WB92)
\bibitem[]{}White R.L., Helfand D.J., Becker R.H., Gregg M.D., 1997,
	ApJ 475, 479

\end{thebibliography}
\end{document}